\documentclass[aps,prl,twocolumn,a4paper,superscriptaddress]{revtex4}

\usepackage[paperwidth=215mm,paperheight=300mm,centering,hmargin=1.6cm,vmargin=2cm]{geometry}
\usepackage{multibbl}
\usepackage{graphicx,amsmath,SIunits}
\usepackage{latexsym,stmaryrd}
\usepackage[sort&compress]{natbib}
\usepackage{color}
\bibpunct{[}{]}{,}{n}{}{,}

\makeatletter
\newcommand*{\citenst}[2][]{%
  \begingroup
  \let\NAT@mbox=\mbox
  \let\@cite\NAT@citenum
  \let\NAT@space\NAT@spacechar
  \let\NAT@super@kern\relax
  \renewcommand\NAT@open{[}%
  \renewcommand\NAT@close{]}%
  \citep{#2}%
  \endgroup
}
\makeatother

\renewcommand{\figurename}{\textbf{Figure}}

\begin{document}

\preprint{APS/123-QED}

\title{Quantifying the robustness of topological slow light.}

\author{Guillermo Arregui}%
\affiliation{Catalan Institute of Nanoscience and Nanotechnology (ICN2), CSIC and BIST, Campus UAB, Bellaterra, 08193 Barcelona, Spain}%

\author{Jordi Gomis-Bresco}
\email{jordi.gomis@icn2.cat}
 \affiliation{Catalan Institute of Nanoscience and Nanotechnology (ICN2), CSIC and BIST, Campus UAB, Bellaterra, 08193 Barcelona, Spain}%

\author{Clivia M. Sotomayor-Torres}%
\affiliation{Catalan Institute of Nanoscience and Nanotechnology (ICN2), CSIC and BIST, Campus UAB, Bellaterra, 08193 Barcelona, Spain}%
\affiliation{ICREA - Instituci\'o Catalana de Recerca i Estudis Avan\c{c}ats, 08010 Barcelona, Spain}

\author{Pedro David Garcia}%
 \email{david.garcia@icn2.cat}
\affiliation{Catalan Institute of Nanoscience and Nanotechnology (ICN2), CSIC and BIST, Campus UAB, Bellaterra, 08193 Barcelona, Spain}%

\date{\today}

\small

\begin{abstract}

Low-dimensional nanostructured materials can guide light propagating with very low group velocity $v_\text{g}$.\ However, this slow light is significantly sensitive to unwanted imperfections in the critical dimensions of the nanostructure.\ The backscattering mean free path, $\xi$, the average ballistic propagation length along the waveguide, quantifies the robustness of slow light against this type of structural disorder.\ This figure of merit determines the crossover between acceptable slow-light transmission affected by minimal scattering losses and a strong backscattering-induced destructive interference when $\xi$ exceeds the waveguide length $L$.\ Here, we calculate the backscattering mean free path for a topological photonic waveguide for a specific and determined amount of disorder and, equally relevant, for a fixed value of the group index $n_\text{g}$ which is the slowdown factor of the group velocity with respect to the speed of light in vacuum.\ These two figures of merit, $\xi$ and $n_\text{g}$, should be taken into account when quantifying the robustness of topological and conventional (non-topological) slow-light transport at the nanoscale.\ Otherwise, any claim on a better performance of topological guided light over conventional one is not justified.

\end{abstract}

 \pacs{(42.70.Qs, 03.65.Vf, 42.25.Dd, 42.25.Fx, 46.65.+g,  42.25.Bs, 78.67.−n)}

\maketitle

Slowing the speed of a light pulse down to human pace (m/s) requires complex interference effects~\cite{Atoms} which manifest as a flat dispersion relation $\nu = \nu (\textbf{k})$, where $\mathbf{k}$  is the conserved wave vector and $\nu(\mathbf{k})$ the frequency.\ The group velocity $v_\text{g}$ of this slow light is determined by the derivative of the flat band and the slowdown factor is given by $n_\text{g} = c/v_\text{g}$, where c is the speed of light in vacuum.\ $n_\text{g}$ is the figure of merit for slow light and it determines the enhancement factor for diverse applications such as optical nonlinearities~\cite{Monat}, optical switching~\cite{Bajcsy}, pulse delay~\cite{Stenner}, quantum optics~\cite{Arcari}, optical storage~\cite{Sayrin} and optical gain~\cite{Ek}.\ Slow light interacts much more efficiently with matter, e.g., atoms~\cite{Atoms}, photon polaritons~\cite{polaritons}, plasmons~\cite{plasmons}, spinors~\cite{spinors}, magnons~\cite{magnons} or excitons~\cite{excitons}.\ A strategy to bring slow light to the nanoscale exploits optical resonances built up by nanostructuring a dielectric or semiconductor with low absorption, such as silicon at telecom wavelengths~\cite{Krauss1,Krauss2,Baba}.\ This potentially disruptive technological platform for enhanced light-matter interaction enables photonic applications ranging from nanolasing~\cite{Mork} and energy harvesting~\cite{Krauss3} to integrated quantum photonics~\cite{Integrated}.\ Flat bands arise naturally in these systems based on the periodic modulation of the refractive index at optical or near infrared wavelengths~\cite{Joannopoulos} for which the group index diverges as $n_\text{g} \propto \left( \partial \nu / \partial k\right)^{-1}$ in the ideal situation.\ However, in real devices there is a limitation to the maximum $n_\text{g}$ achievable due to slight deviations of the fabricated parameters compared to the designed values.\ Even fluctuations in the nm-range~\cite{PD1} give rise to backscattering of the guided light, inducing a strong interference~\cite{Topolancik}, a photonic manifestation of Anderson localization in low dimensions~\cite{Anderson}.\ Imperfection limits slow light in conventional photonic waveguides to maximum values around $n_\text{g} \approx 100$ for very short waveguides with lengths $L \simeq 5\,\mu\text{m}$, much lower than the $n_\text{g}$ values observed in atomic systems~\cite{Atoms} but still sufficiently large to explore weak light-matter interaction leading to cavity-quantum electrodynamic phenomena~\cite{Arcari}.\ In this Letter, we confirm that this limitation may be overcome by exploiting photonic topological effects that purely arise from engineering the lattice geometry.

\begin{figure}[t!]
  \includegraphics[width=\columnwidth]{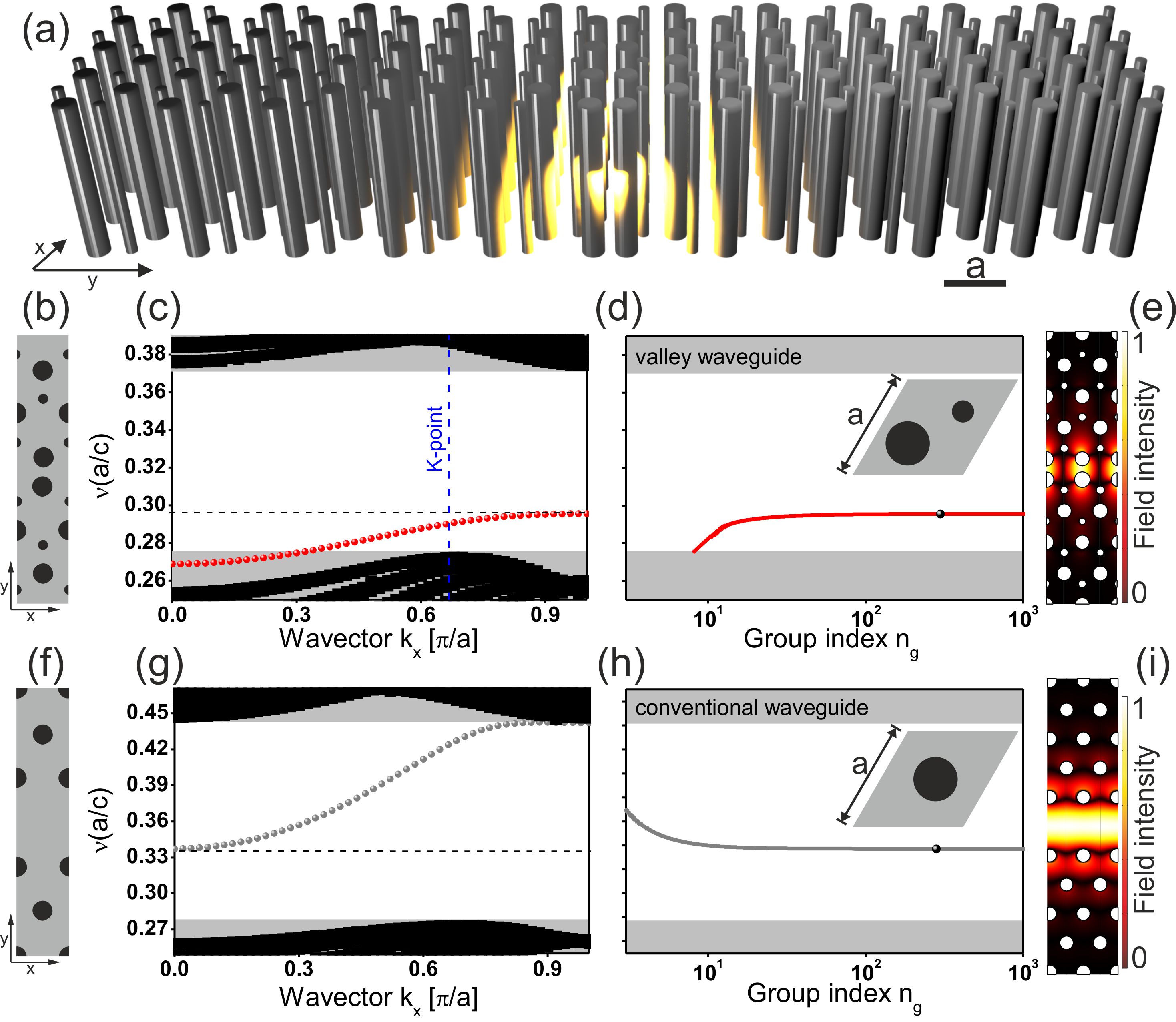}
    \caption{ \label{1} \textbf{Slow light in topological waveguides.}
     (\textbf{a}) Illustration of a valley topological waveguide formed at the interface of two valley crystals with different topological invariants as proposed in Ref.~\cite{valley}.\ (\textbf{b}) Distribution of the dielectric function in the structure: black corresponds to silicon and gray to air.\ Each valley crystal is formed as a triangular periodic lattice with a unit cell composed by two pillars with different diameters to break the spatial inversion symmetry.\ (\textbf{c}) Dispersion relation, $\nu = \nu (k)$, and (\textbf{d}) calculated group index of the interface topological edge state.\ The unit cell is detiled in the inset.\ For reference, we design a conventional photonic waveguide obtained by leaving a row of pillars from a triangular lattice (\textbf{f}) with a desertion relation and a group index plotted in (\textbf{g} and (\textbf{h})), respectively.\ The black point in the group index curves denotes the frequency at which each waveguide has a group index of $n_\text{g} \approx 300$.\ The electromagnetic field intensities calculated at these frequencies and for perfect (non-disordered) waveguides are plotted in (\textbf{e}) and (\textbf{i}).}
\end{figure}

\begin{figure*}[t!]
  \includegraphics[width=\textwidth]{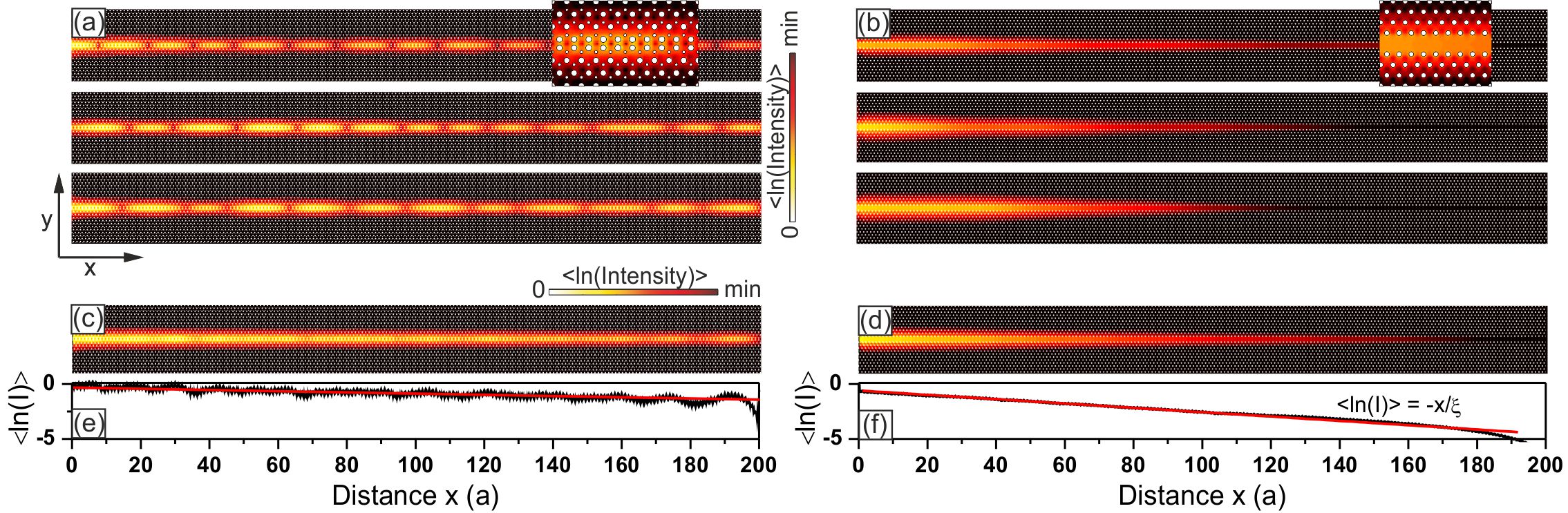}
    \caption{ \label{2} \textbf{Backscattering in topological slow-light waveguides.}
    (\textbf{a}) Normalized electromagnetic-field intensity excited by a dipole emitter positioned on the left side of a disordered valley-Hall waveguide at a frequency $\nu = 0.2955(c/a)$, where $a$ is the lattice constant and c is the speed of light in vacuum.\ Three different configurations of randomized positions of the pillars are plotted with a standard deviation of $\sigma = 0.001a$.\ (\textbf{b}) Calculations for a conventional photonic waveguide where the dipole emitter is excited at a frequency $\nu = 0.3372(c/a)$.\ In (\textbf{a}) and (\textbf{b}), the dipole emission frequency corresponds to a group index of $n_\text{g} = 300$.\ (\textbf{c}) and (\textbf{d}) show the ensemble-averaged electromagnetic field intensity excited by the dipole source at the same frequencies of \textbf{a} and \textbf{b}.\ For this calculation, over twenty configurations of positional disorder of the pillars with $\sigma = 0.001a$ were averaged.\ (\textbf{e}) and (\textbf{f}) illustrate the calculated ensemble-averaged electromagnetic field intensity profile along the topological and conventional waveguide axis, respectively.\ The localization length is extracted from the inverse of the exponential decay slope.}
\end{figure*}

Topological photonics has emerged very recently as a competitive approach for robust light transport~\cite{TopologicalPhotonics}, something extremely appealing for technological applications.\ There is a broad spectrum of possible implementations of photonic topological insulators which translate concepts from solid state to optical systems.\ The most robust one implements time-reversal symmetry broken edge states~\cite{AQHE} for which the time-reversed state at -\textbf{k} does not exist, thus preventing direct elastic backscattering of the propagating mode.\ However, to implement this approach in photonics one requires strong magnetic effects which at visible and near-infrared wavelengths are very weak and challenging to implement~\cite{Marin}.\ In time-invariant topological insulators based on the quantum-spin Hall effect~\cite{QSHE} and the valley-Hall effect~\cite{valley}, topology emanates from the breaking of particular spatial symmetries.\ In such implementations, reciprocity imposes the existence of the counter propagating mode at -\textbf{k}.\ This time-reversed edge state carries the opposite value of a binary degree of freedom that plays the role of a pseudo-spin~\cite{QSHE}.\ In this case, the key open question is whether backscattering is reduced and the answer will depend on whether the existing structural disorder preserves the pseudo-spin value or not.\ Another rather interesting question is whether the symmetries of the topological system which determine its topological invariants are preserved under different types of perturbations.\ Recent ground breaking experiments reported robustness in terms of a certain lack of structural back-reflection when precisely-shaped local defects were introduced in different topological waveguides~\cite{valley_exp}.\ However, this claimed robustness still needs to be systematically quantified against disorder, which is crucial to compare properly these waveguides to the state-of-the art conventional (non-topological) ones.\ Here, we engineer slow light in a valley-Hall waveguide~\cite{valley} and we calculate its backscattering length, $\xi$, versus disorder and $n_\text{g}$.\ Finally, we compare the results to those of a conventional photonic waveguide.

For our analysis, we focus on the parity-symmetry breaking valley-Hall approach~\cite{valley}.\ When applied to photonic slabs, the topological edge states at an interface between valley Hall crystals of opposite K-valley pseudospin lay below the light line of the slab and are decoupled from the radiation continuum.\ In principle, these edge states have no intrinsic out-of plane losses and only fabrication imperfection can induce coupling to radiating modes.\ Even in that situation, in plane back-scattering is largely the dominant loss mechanism at large values of $n_\text{g}$~\cite{Mazoyer} so it is enough to consider the system as two dimensional to capture the physics of slow light backscattering, something not possible with other implementations of topological photonics~\cite{Sauer}.\ Therefore, we implement two dimensional simulations instead of the more computationally expensive three-dimensional slab.\ Fig.~\ref{1}(a) displays an illustration of a fully two-dimensional valley-Hall waveguide formed at the interface of two topologically different valley crystals with bandgaps induced by breaking spatial inversion symmetry.\ The valley crystals are created with a triangular lattice where the unit cell is formed by two circular silicon pillars surrounded by air with different diameters $\text{d}_1 = 0.4a$ and $\text{d}_2 = 0.2a$, where $a$ is the lattice constant.\ The section of the waveguide is plotted in Fig.~\ref{1}(b).\ The dielectric constants for silicon and air are taken here as $\varepsilon_{Si}=12$ and $\varepsilon_{air}=1$, respectively.\ For $\text{d}_1 = \text{d}_2$, the system preserves the $C_{6v}$ symmetry and supports a symmetry-protected gapless band structure between the first and second lowest energy bands for transverse magnetic polarized light.\ When the pillars have a different diameter, the system breaks the spatial inversion symmetry opening a bandgap between these two bands~\cite{valley}.\ Fig.~\ref{1}(c) shows the calculated dispersion relation of the waveguide and in Fig.~\ref{1}(d) we plot the group index of the topological edge mode with vanishing group velocity at the cutoff frequency $\nu = 0.2955(c/a)$, which corresponds to $\nu = 177\,\text{THz}$ for $a = 500\,\text{nm}$.\ For reference, we use a standard photonic crystal waveguide obtained by leaving out a row of pillars in a triangular lattice of silicon pillars surrounded by air.\ The diameter of the pillar in this case is set to $\text{d} = 0.4a$ with the same lattice unit as the topological waveguide, as shown in Fig.~\ref{1}(f).\ The guided mode of this conventional waveguide presents an ideally vanishing group velocity, or diverging group index, when approaching the cutoff frequency of the waveguide at $\nu = 0.3372(c/a)$, as shown in Fig.~\ref{1}(g) and (h).\ For the sake of clarity, we bring the attention to the fact that the cutoff of the topological and conventional waveguide lie on the X and $\Gamma$ point, respectively.\ The ideal spatial field-intensity distributions in both waveguides, i.e., in absence of any perturbation, and at frequencies corresponding to $n_\text{g} = 300$ in both cases are plotted in Figs.~\ref{1}(e) and (i) for reference, evidencing a similar level of light confinement.\ Using silicon pillars surrounded by air instead of the usual air holes in silicon enables us to flatten the dispersion relation of both topological and conventional guided modes successfully without the need of local perturbations as in Ref.~\cite{EngineeredSlowLight} or progressive interfaces as done in Ref.~\cite{BZWinding}.\ The parameters to obtain flat bands in the valley-Hall waveguide are detailed in the supplementary material.

The backscattering length, $\xi$, is the average ballistic propagation distance along the waveguide in absence of any other major loss mechanism~\cite{Mazoyer}.\ A slow-light waveguide becomes virtually useless when $L \gg \xi$.\ Interestingly, $\xi$ is governed by the density of optical states of the waveguide, at least for a weak perturbation of the wavegudie~\cite{Mazoyer,PD3}, as $\xi \propto \text{DoS}^{-1} = \left( \partial \nu / \partial k\right)$.\ Intuitively, a larger density of optical states induces a larger probability of scattering, thus reducing the value of $\xi$.\ As $n_\text{g} \propto \left( \partial \nu / \partial k\right)^{-1} = \text{DoS}$, both the group index and the backscattering length are intrinsically linked to each other via the DoS, at least in conventional photonic crystal waveguides~\cite{PD4}.\ Despite substantial theoretical work on $\xi$ in non-topological electronic~\cite{Garcia-Martin, Froufe, MacKinnon} and photonic transport~\cite{Mazoyer, Savona, PD4}, this parameter has only been explored recently in topological waveguides~\cite{Fleury} although ignoring $n_\text{g}$.\ It is important to remark the fact that the backscattering length is a universal parameter in one-dimensional transport~\cite{Sheng, Smolka} and any mesoscopic-transport observable in these systems depends only on $\xi$, regardless of the microscopic details of the medium considered.\ Therefore, two waveguides with a different nanostructured pattern but with the same $\xi$ share the same universal mesoscopic-transport properties, regardless of it being silicon pillars surrounded by air or air inclusions in silicon.\ We calculate $\xi$ in perturbed topological and conventional photonic crystal waveguides as:

\begin{equation}
 - \frac{x}{\xi(\nu)} = \langle \ln [I(\nu)] \rangle
 \label{xi}
\end{equation}
where $I$ is the finite-element solution of the electromagnetic field intensity emitted by a dipole at frequency $\nu$, $x$ is the distance from the dipole position along the waveguide and the brackets indicate the statistical ensemble average over different configurations of positional disorder.\ Our simulation domain has a length $L = 200a$ in the $x$ direction, eleven unit cells on each side of the waveguide in the $y$ direction and it is all surrounded by perfectly-matched layers to mimic an open system.\ To simulate the effect of fabrication imperfection, we randomize the position of the pillars around their ideal value according to a normal distribution which standard deviation $\sigma$ is our measure of disorder (more details in the supplementary material).

The electromagnetic field intensity excited by a dipole source oscillating at the leftmost edge of the waveguide is shown in Fig.~\ref{2}.\ The oscillation frequency is chosen such that both the topological (left) and the conventional (right) waveguide would enable slow light transmission with $n_\text{g} = 300$ in the absence of any imperfection.\ The figure shows calculations corresponding to varying configurations of positional disorder for which the positions of the pillars are randomized with a fixed $\sigma = 0.001a$.\ As shown in Fig.~\ref{2}(a), the excited Bloch modes in the topological waveguide are just slightly perturbed, which reveals a rather weak backscattering for this large group index value.\ However, strong backscattering interference prevents light transport in the conventional waveguide, as revealed by the different examples plotted in Fig.~\ref{2}(b).\ As $\xi$ is a statistical parameter, this requires an ensemble-average calculation of many (ideally all) different disorder configurations.\ This ensemble average is not easily accessible in experiments because it is difficult to discriminate backscattering from other loss sources, particularly absorption~\cite{Red}.\ In numerical experiments, as the one performed here, $\xi$ is easily obtained by computing the position-dependent field excited by the dipole source in the conditions described above for several structural configurations with the same nominal amount and type of disorder.\ Figs.~\ref{2}(c) and (d) show the intensity pattern after ensemble-averaging the electromagnetic field excited by this emitter over twenty different realizations, the envelope of which decays exponentially from the position of the source with a sufficiently well-averaged slope, as plotted along the waveguide axis in Figs.~\ref{2}(e) and (f) for $n_\text{g} = 300$.\

\begin{figure}
  \includegraphics[width=\columnwidth]{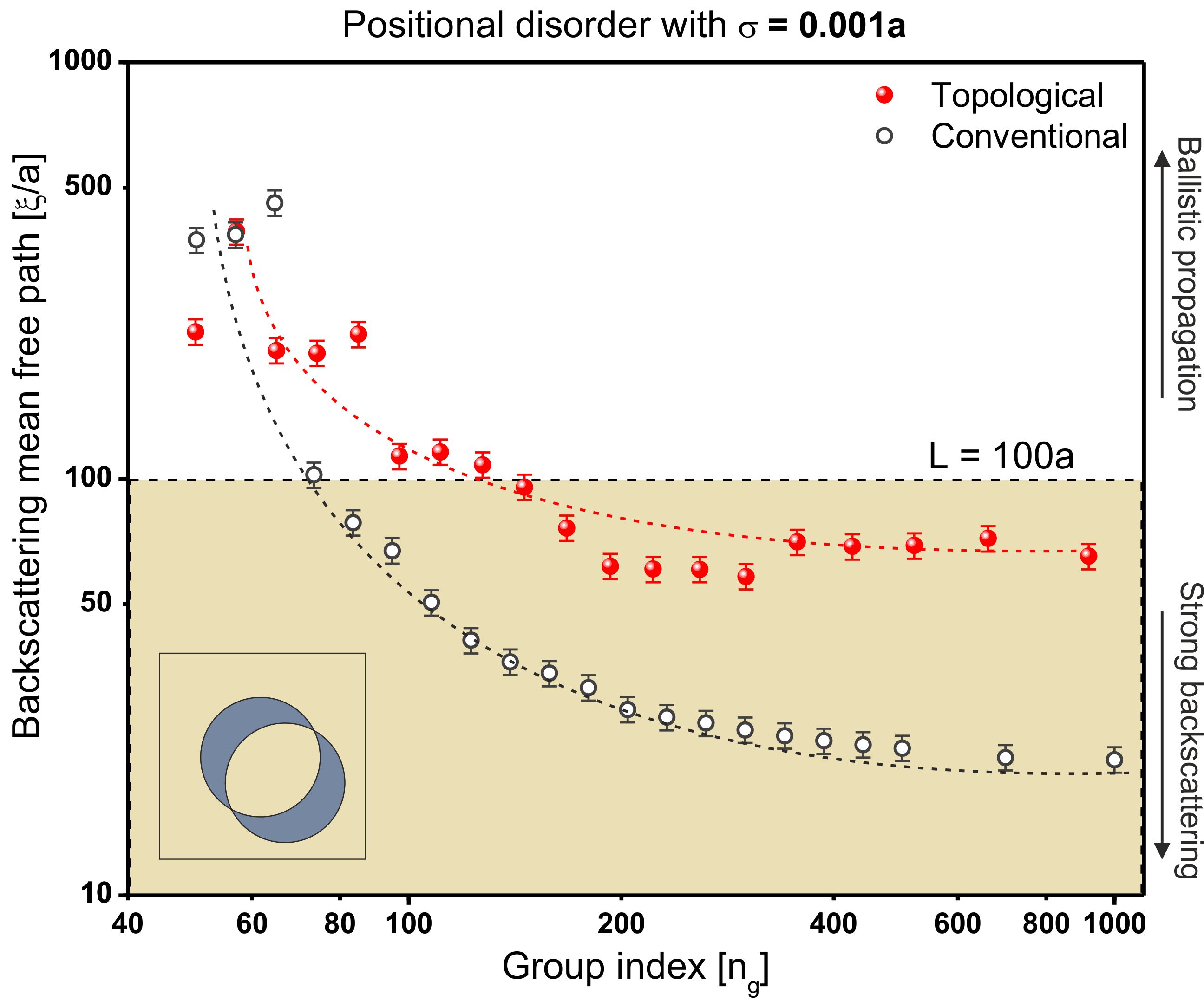}
    \caption{ \label{3} \textbf{Backscattering length versus group index in topological waveguides.}
    (\textbf{a}) Backscattering length calculated for different frequencies corresponding to different values of the group index in a valley and a conventional photonic crystal waveguide.\ The positions of the pillars have been randomized around their ideal values with a standard deviations of $\sigma = 0.001a$.\ The shaded area indicates the crossover between ballistic transport and strong backscattering for a waveguide of length $L = 100a$.\ The error bars are a conservative estimate of the fitting error to the decay slope as plotted in Figs.~\ref{2}(e) and (f).\ The lines are guides to the eye.}
\end{figure}

\begin{figure}
  \includegraphics[width=\columnwidth]{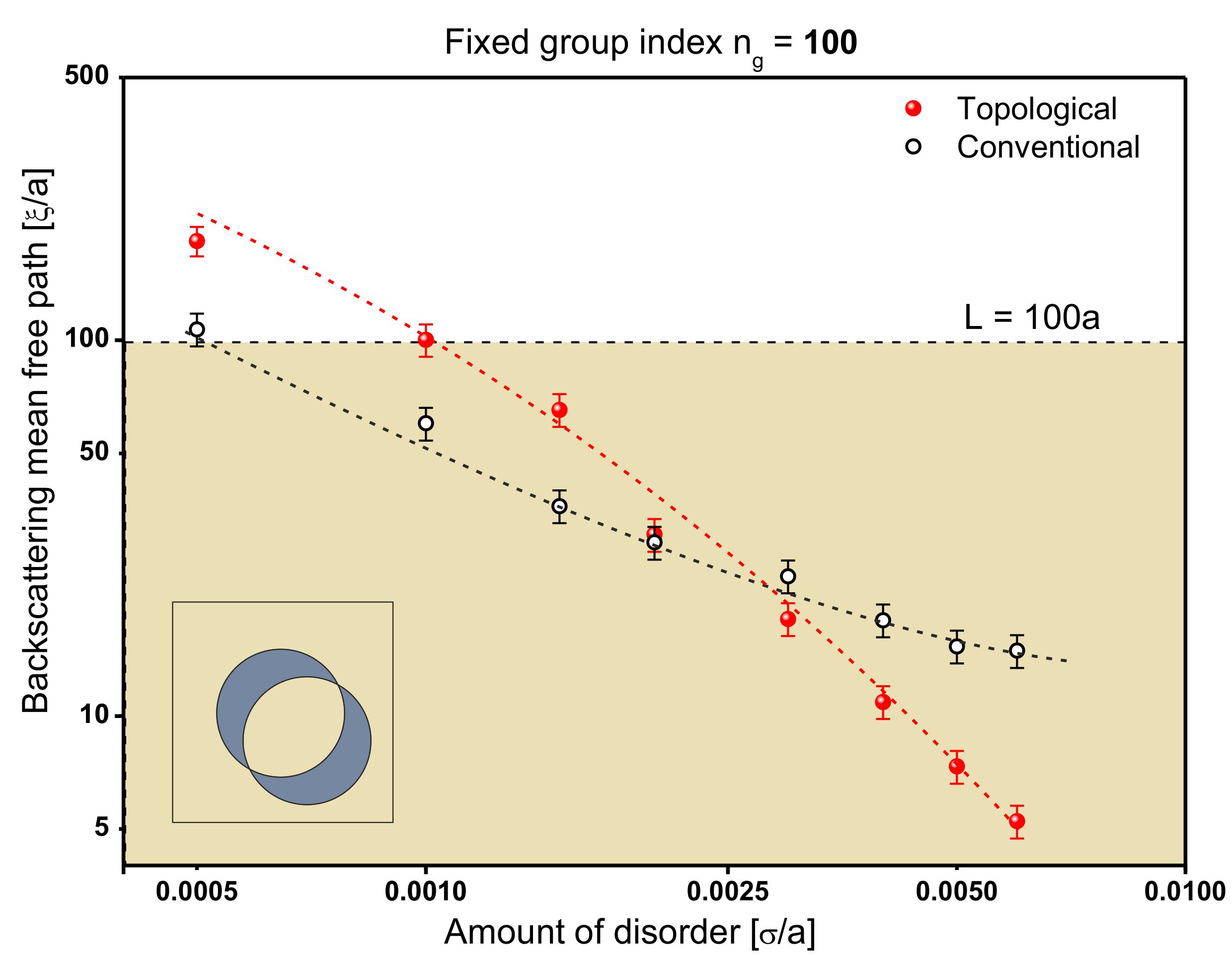}
    \caption{ \label{4} \textbf{Backscattering length versus disorder in topological waveguides.}
    Backscattering length calculated by randomizing the position of the pillars by an increasing amount for a fixed group index value $n_\text{g} = 100$.\ The shaded area indicates the crossover between ballistic transport and strong backscattering for a waveguide of length $L = 100a$.\ The lines are guides to the eye.}
\end{figure}

The backscattering length $\xi$ calculated for the topological and conventional waveguides for different $n_\text{g}$ is plotted in Fig.~\ref{3}.\ $\xi$ is given for a fixed amount of positional disorder $\sigma = 0.001a$, which corresponds to fluctuations of $\approx 0.5\,\text{nm}$ for $a = 500\,\text{nm}$.\ This is a realistic measure of the residual imperfection resulting from a state-of-the art fabrication process~\cite{PD1, Galli}.\ The figure reveals the fact that, at low disorder levels, the topological waveguide considered here and based on a geometry-engineered topological phase is more robust than standard conventional slow-light waveguides when the full phase is randomized, thus mimicking the effect of imperfection in real systems.\ As shown in Fig.~\ref{3}, topological waveguides suffer much less backscattering than conventional ones with a backscattering length comparable to the waveguide length for large group index values.\ Even at large values of the group index, $n_\text{g} \simeq 1000$, the interface edge mode mode of the topological waveguide is slightly perturbed when compared to the strong backscattering suffered by the conventional one, as shown in the different configurations plotted in Fig.~\ref{2}(a).\ The backscattering mean free path for various amounts of disorder and a fixed $n_\text{g} = 100$ is plotted in Fig.~\ref{4}.\ Below a critical level of disorder, $\sigma_\text{c} \sim 0.002a$, the topological waveguide outperforms the conventional one.\ Above this level, the probability of a pseudo-spin flip upon scattering of the Bloch mode increases, leading to decrasing $\xi$.\ This behaviour was confirmed for various values of the group index at lower disorder levels.\ We attribute this to a topological protection which disappears at a sufficiently high degree of disorder, in this work at $\sigma_\text{c} = 0.002a$, as shown in Fig.~\ref{4}.\ Above $\sigma_\text{c}$, the topological waveguide suffers an even stronger backscattering than the conventional one.\ We understand this behaviour above $\sigma_\text{c}$ as due to the effective mass of the propagating topological Bloch mode in absence of any perturbation~\cite{PD4}.\ A larger effective mass (flatter bands) induces stronger backscattering~\cite{Savona,Fagiani}, which is what we observe in Fig.~\ref{4} above the critical level of disorder.\ In our case, the effective mass of the topological waveguide is almost an order of magnitude larger than the conventional one, which underlines even more the potential of the topological protection observed below $\sigma_\text{c}$ and for very large values of $n_\text{g}$, the main result of this Letter.

In conclusion, the robustness of diverse designs of topological photonic waveguides based on pseudo quantum-spin Hall and valley-Hall effect against structural back-reflection has been claimed in recent experiments~\cite{AQHE_exp, QSHE_exp, valley_exp, Shalaev}, where the sole inclusion of local defects in perfect topological phases was taken into account.\ However, one of the key elements to quantify the robustness of topological photonic platforms is to randomize the full topological phase to mimic real imperfection due to the fabrication process, which is the real limitation in slow light transport in state-of-the art photonic crystal waveguides.\ We have shown how to quantify the robustness of a topological edge state against white noise on its structural parameters.\ Calculating the backscattering length linked to the group index enables us to do that, as these two parameters are related to each other through the density of optical states.\ We conclude that current proposals of topological photonic phases relying on the breaking of different parity symmetries as the valley-Hall effect are quantitatively, by almost five times, more robust than standard conventional waveguides with small disorder levels, although this protection is lost at higher imperfection amount.\ We expect this protection to manifest at even larger disorder levels if we take into account the much larger effective masses of the ideal topological edge modes, which plays a dominant role for light localization at frequencies close to the cutoff one.\ The analysis carried out here is based on a particular system of silicon pillars surrounded by air but the approach is completely generic and can be implemented for any other single-mode waveguide with arbitrary design.\ Nonetheless, our results point to the importance of the group index $n_\text{g}$ for a quantitative comparison of the robustness of light transport.\ The large values of the group index calculated here, $n_\text{g} \simeq 1000$, affected by a relatively weak backscattering provide a promising platform for highly efficient strong light-matter interaction~\cite{Goban} where photon transmission over hundreds of microns is relatively backscattering-free.\ Our analysis, in combination with the calculation of the topological invariants applying topological quantum chemistry~\cite{Aitzol} provides a necessary tool to test the validity of any translation of topological effects from solid-state physics to photonics.\ Future work to evaluate topological invariants of different topological implementations will provide additional insight on the relationship between the backscattering length evolution with disorder level and the protection granted by topology.

\section*{Acknowledgments}

This work was supported by the European Commission H2020 FET Proactive project TOCHA (No. 824140).\ The authors acknowledge funding by the Severo Ochoa program from Spanish MINECO (Grant No. SEV-2019- 0706) and the Plan Nacional RTI2018-093921-A-C44 (SMOOTH) and CERCA Programme/Generalitat de Catalunya.\ G.A. was supported by a H2020 MSC cofund action; a BIST Ph.D. fellowship and P.D.G. by a Ramon y Cajal Fellowship No. RyC-2015-18124.


\begin{widetext}

\newpage

\renewcommand{\figurename}{\textbf{Supplementary Figure}}
\makeatletter
\renewcommand{\thefigure}{S\@arabic\c@figure}
\makeatother
\renewcommand\theequation{S\arabic{equation}}
\renewcommand\thetable{S\arabic{table}}
\renewcommand{\bibname}{References}

\section{Supplementary information}

\setcounter{figure}{0}

\section{valley-Hall topological photonics.}

In time-invariant solid-state systems, topological effects are attributed to mechanisms that break the parity symmetry.\ In the quantum-spin Hall effect, the spin-orbit coupling plays the role of the external magnetic field to induce the topological protection.\ In that case, the spin up and down of the electron exhibit chiral and anti-chiral integer quantum Hall effect, respectively.\ To translate these mechanisms from solid state to photonics, we need to find degrees of freedom that mimic that of the electron spin in order to build up a pseudospin for light~\cite{WuHu}.\ Alternatively, one can implement a valley-Hall effect~\cite{Benakker} which exploits the angular rotation of the electron wave function in the $K$ or $K'$ valleys of the band structure generating an intrinsic magnetic moment~\cite{magnetic} analogous to that produced by the electron spin.\

The strategy to implement a valley-Hall topological photonic phase exploits just a single polarization (transverse magnetic in our case) in a lattice where the inversion symmetry is broken in the unit cell.\ This is fairly more simple than current implementations of the quantum-spin Hall effect for photonics, which require the folding of the dispersion relation.\ The idea behind this implementation~\cite{valley} is to create an interface between two different valley-Hall phases with different symmetry-breaking geometries which supports highly-confined topological edge-state guided modes.\ All the details can be found in the seminal paper written by Ma and Shvets~\cite{valley}.

\section{Finite element calculations and disorder.}

\begin{figure}[b!]
  \includegraphics[width=\columnwidth]{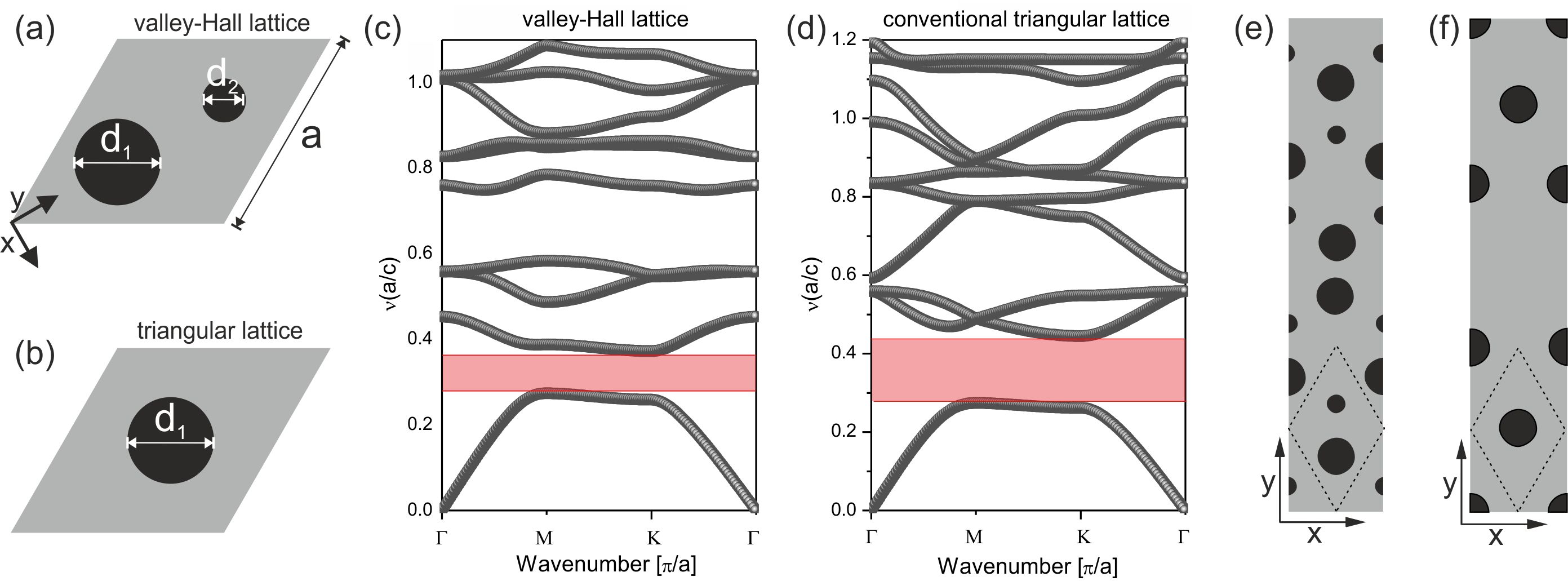}
    \caption{ \label{S1} \textbf{Primitive lattices and dispersion relations.} (a) Primitive lattice of a valley-Hall crystal formed by two pillars of silicon (black) surrounded by air (gray) with different diameters $d_1 =0.4a$ and $d_2 =0.2a$).\ (a) Primitive cell of a conventional triangular crystal formed by a silicon pillar surrounded by air with a defeater $d_1 =0.4a$.\ (c) and (d) display the eigenmode dispersion relation for the transverse magnetic field for a triangular lattice with the primitive cell plotted in (a) and (b), respectively.}
\end{figure}

Most of recent experiments in topological photonics explore slab geometries, i.e., material with a subwavelength thickness.\ In the particular case of the valley-Hall topological approach~\cite{valley}, the edge states appear below the light line, and are therefore perfectly guided.\ Disorder is a coupling mechanism between these modes and the radiation continuum, however, in the slow-light region of a band, in plane scattering dominates largely over out-of plane radiation losses~\cite{Mazoyer}.\ Based on this, we neglect the out-of-plane dimension and implement a two-dimensional geometry, capturing the physics of the problem at a reduced numerical cost.\ We perform a finite-element calculation using commercially available software to find the solution of the Maxwell equations to this two-dimensional problem.\ To obtain the variation of the backscattering length in a wide range of group index values, we engineer slow light in a valley-Hall waveguide by creating a lattice formed of silicon pillars surrounded by air.\ Our design is a valley-Hall waveguide based on a primitive cell of triangular symmetry of size $a$ composed by two dielectric silicon pillars with a dielectric constant $\varepsilon_{Si}=12$ surrounded by air $\varepsilon_{air}=1$, as detailed in Fig.~\ref{S1}(a).\ The two pillars of the primitive cell have different diameters ($d_1 =0.4a$ and $d_2 =0.2a$) to break the inversion symmetry.\ The coordinates of the center of the small and large pillar within the primitive cell are $\{0, 4 \sqrt 3a/6 \}$ and $\{0, \sqrt 3a/3 \}$, respectively, taking the down-right corner as origin and the $x$ and $y$ coordinates as indicated in Fig.~\ref{S1}(a).\ For this configuration, the transverse magnetic field presents a frequency band gap between $0.2740$ and $0.3727$ in normalized units, i.e., in units of $a/c$.\ The gap is shaded in red in Fig.~\ref{S1}(c).\ Finally, we build the valley-Hall waveguide by interfacing two valley-Hall crystals with the same geometry and opposed symmetry, as displayed in Fig.~\ref{S1}(e).\ In the calculations performed here, the pillars with larger diameter are closer to each other and their centers separated by a distance equal to $\sqrt 3/3a$.\ In this configuration, we obtain a single topological edge mode and presents the cutoff at the X symmetry point.\ We have attempted to engineer slow light in a configuration of air inclusions in silicon without so far success.

For reference, we design a (conventional) non-topological photonic waveguide with a similar configuration based on a primitive cell of triangular symmetry of size $a$ composed by a single dielectric pillar ($\varepsilon_{Si}=12$) of diameter $d=0.4a$ surrounded by air ($\varepsilon_{air}=1$), as shown in Fig.~\ref{S1}(b).\ The transverse magnetic field presents a frequency band gap for this lattice between $0.2740$ and $0.4449$ in normalized units, as plotted in Fig.~\ref{S1}(d).\ The conventional waveguide is built by leaving out a row of pillars creating a line defect in the $x$ direction, as detailed in Fig.~\ref{S1}(f).\

We introduce disorder in both waveguides by displacing the position of all the pillars independently a random distance from their position in the perfect lattice, $\Delta \text{\textbf{r}}$.\ This distance is normally distributed with a standard deviation $\sigma = \sqrt{\langle \Delta \text{\textbf{r}}^2\rangle}$ and $\langle \Delta \text{\textbf{r}} \rangle = 0$, where brackets indicate the ensemble average over all configurations of random fluctuations.

For the calculation of the unperturbed waveguide, we simulate a numerical domain formed by a single unit cell in the $x$ (propagation) direction and $22$ cells in the $y$ direction.\ As boundaries of our simulation domain we terminate the waveguide with Bloch periodic conditions in the $x$ direction and perfectly-matched layers in the $y$ direction.\ For the calculation of the disordered waveguides, we expand the numerical domain in the $x$ (propagation) direction to 200 unit cells.\ To simulate an open system where light can leak out, we introduce perfectly-matched layers surrounding the whole computational domain.\ This mimics open boundaries for which electromagnetic waves are transmitted with minimal reflection.\ Despite the presence of such perfect-matched layers, slight reflections occur at the interfaces, which results in the generation of extremely broad Fabry-P\'{e}rot resonances whenever the system size $L<\xi$.\ In all our calculations, we implement a mesh with a ratio 150 elements per $a^2$ area.

\section{Convergence of the results}

\begin{figure}[b!]
  \includegraphics[width=15cm]{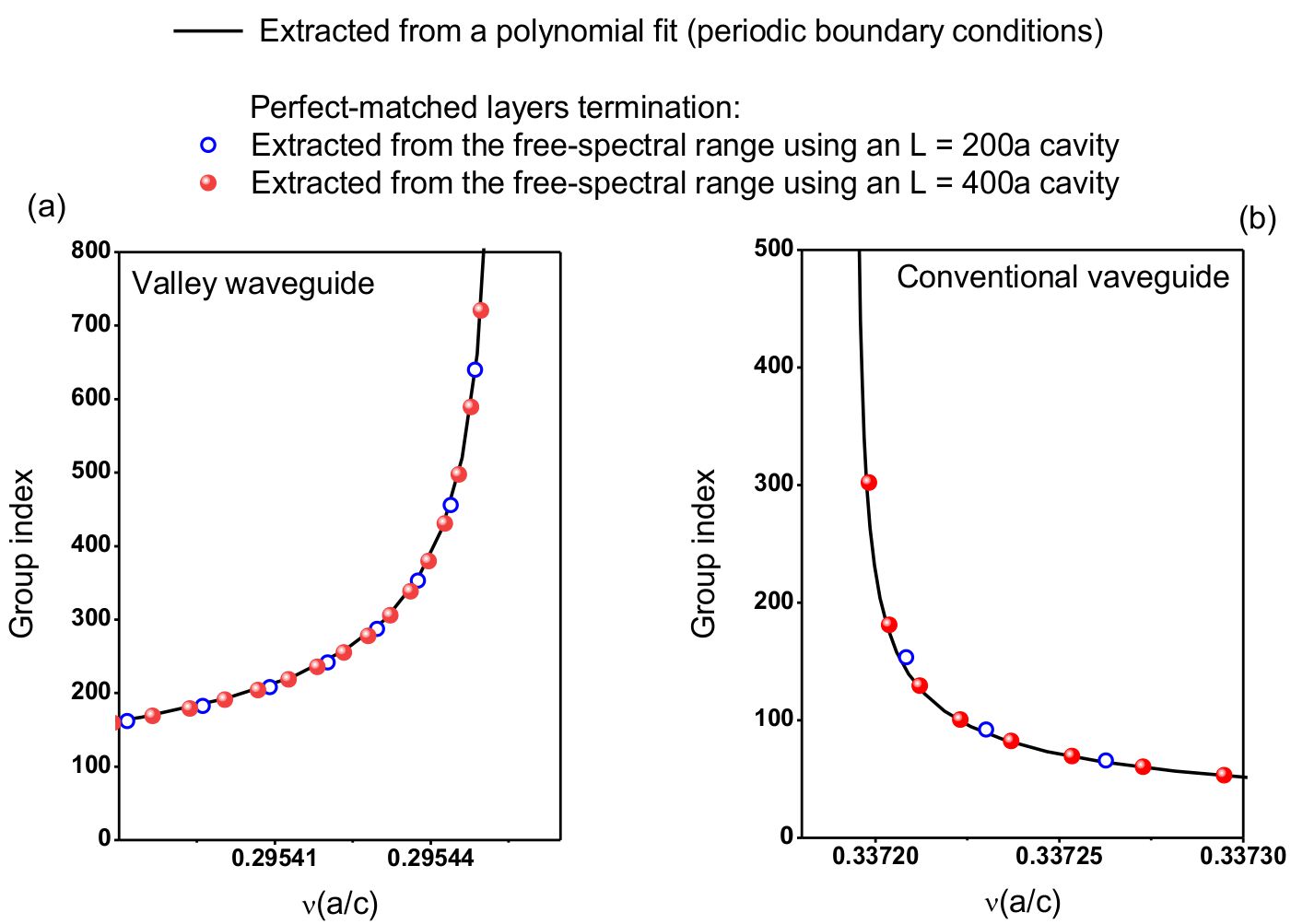}
    \caption{ \label{S2} \textbf{Group index calculated for a valley-Hall (a) and a conventional (b) waveguide.}
    We use two different methods to obtain these values with two different boundary conditions.\ First, by fitting the calculated eigenmode solution for a unit cell with periodic boundary conditions six order polynomial.\ We obtain the group index by simply deriving its value it from the fitting function (black solid line).\ Alternatively, we simulate a full perfect waveguide with two different lengths and perfect-matched layers at both terminations.\ Residual reflection for these terminations induce Fabry-P\'{e}rot resonances.\ We obtain the value of the group index directly rom the free-spectral range of these resonances, plotted in empty and solid dots for $L=200a$ and $400a$, respectively.}
\end{figure}

To test the convergence of our results, we first test the robustness of the calculated waveguide eigenmode frequencies for an increasing number of elements in the mesh until we observe a saturation in their values.\ We start with 18 elements per unit area ($a^2$) and we finish with 200 elements per unit area.\ We observe that the values of the calculated frequencies do not change significantly when we have 150 or more elements per unit area.\ We fix the mesh with a ratio of elements per surface of the waveguides at 150 elements per $a^2$ area for all our calculations and, in particular, for all the disordered configurations.\

In addition, we use two methods to evaluate the group index and to further test the convergence of our results, in particular the value of the group index.\ We do perform this security check to ensure that we have the same values of the group index for the eigenmode calculation with periodic boundary conditions and for an extended numerical domain calculation terminated by perfect-matched layers.\ In one case, we calculate the dispersion relation $\nu(k)$ of a single perfect unit cell with periodic boundary conditions.\ Then, we obtain the $n_\text{g}$ by fitting the calculated dispersion to a sixth order polynomial $p(k)$.\ Finally, we derive the value of the group index $n_\text{g}$ from this polinomial function as:
\begin{equation}
n_g=\frac{c}{v_g}=\frac{1}{c}\left[ \frac{\partial p(k)}{\partial k } \right]^{-1}
\end{equation}

The values of the group index obtained with this method for the valley-Hall and the conventional waveguide are plotted as a solid black line in Figs.~\ref{S2}(a) and (b), respectively.

The alternative method relies on calculating the eigenmodes of unperturbed realizations with the same calculation configuration that we use to extract the backscattering mean free path.\ We simulate waveguides with two different lengths $L=200a, 400a$ and we terminate them in the $x$ (propagation) direction by perfectly-matched layers.\ As these layers are not $100\,\%$ reflectionless and allow some minimal but sufficient reflection, we do observe clear Fabry-P\'{e}rot modes corresponding to the modes of an effectively closed cavity with same length.\ We calculate the value of the group index from the free-spectral range between two Fabry-P\'{e}rot modes, $\Delta\nu_{\text{FSR}}=\nu_{n+1}-\nu_{n}$, as:
\begin{equation}
n_g=\frac{c}{\Delta\nu_{\text{FSR}}L}
\end{equation}
where we approximate the frequency value at which $n_\text{g}$ is obtained simply by the mean value $\nu_{n}=\frac{\nu_{n}+\nu_{n+1}}{2}$ between the resonances.\ The group index calculated using this method for the two different waveguide lengths, $L=200a$ and $400a$, is plotted in Fig.~\ref{S2}(a) and (b) as empty and solid dots, respectively.\ The good correspondence between the values of the $n_\text{g}$ extracted using both methods is a strong convergence test of the meshes employed throughout the whole manuscript.\ More importantly, it ensures us that the calculated values of the group index used to characterize the backscattering mean free path in the main text are correct and robust.

\section{Determination of the effective photon mass and group index}

\begin{figure}[b!]
  \includegraphics[width=\columnwidth]{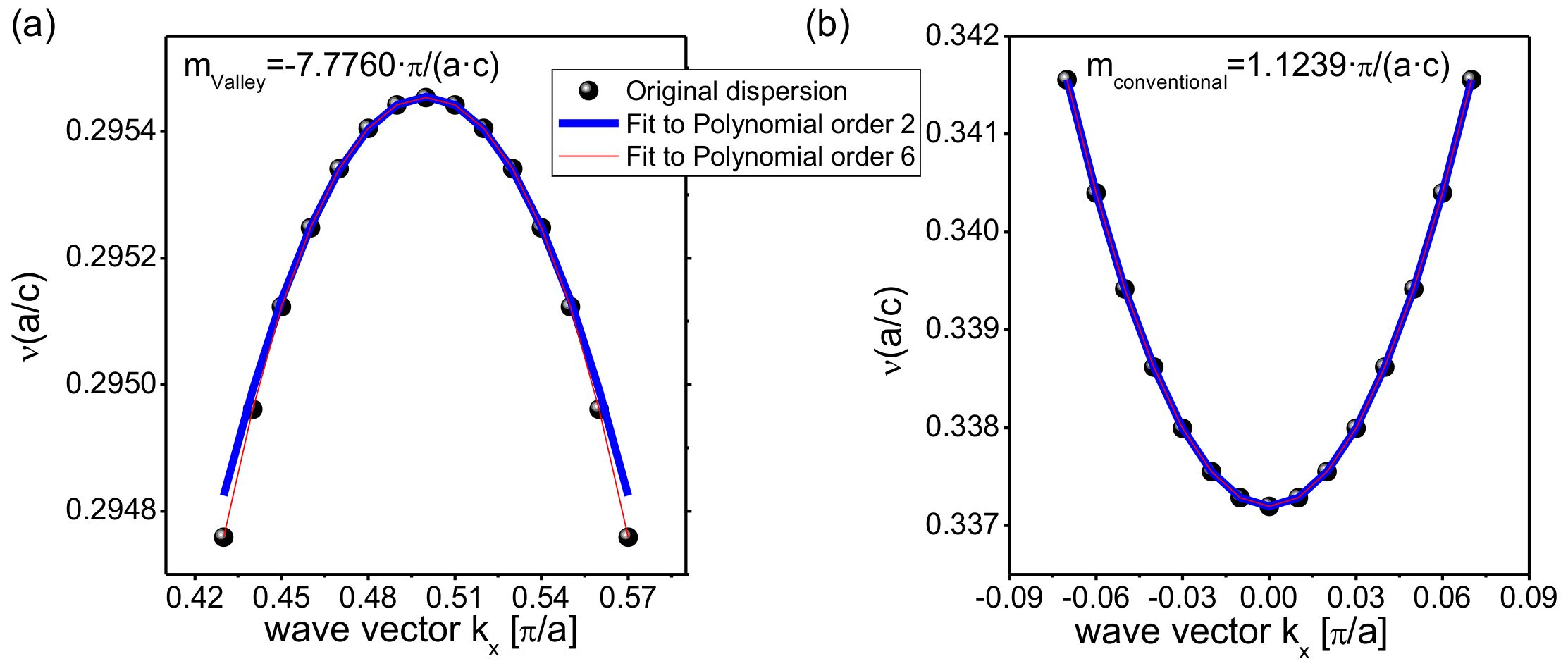}
    \caption{ \label{S3} \textbf{Zoom of the dispersion around the slow light maxima.}
    (a) for the valley-Hall and (b) for the conventional waveguide.\ The blue solid lines correspond to second order fits used to determine the effective photon mass as described in the text, while red lines correspond to a higher order polynomial fit used to extract $n_g$}
\end{figure}

The effective mass of the Bloch mode is defined as $m_\text{eff}=\left[\frac{\partial^2 \nu}{\partial k^2}\right]^{-1}$, where $\nu$ and $k$ are the frequency and wave vector.\ By fitting the band dispersion close to the band gap edge to a second order polynomial, we can extract $m_\text{eff}$ from the second order fitting coefficient.\ In the supplementary Figs.~\ref{S3}(a) and (b), we obtain the effective mass corresponding to the propagating Bloch mode of interest for the conventional and valley-Hall waveguides, respectively.\ In particular, we obtain $m_\text{Valley}= -7.7760\pi/(ac)$ and $m_\text{conventional}= 1.1239\pi/(ac)$.\ To obtain a good fit, we limit the data to a small range around the symmetry point.\ In the case of the conventional waveguide around the $\Gamma$ point, $k_x=\left[-0.1,0.1\right]$ and in the case of the valley-Hall waveguide around the $M$ point, $k_x=\left[0.9,1.1\right]$.\ From our calculations, we obtain an order of magnitude larger effective mass for the valley-Hall edge state when compared to the Bloch mode of the conventional waveguide.

The effective mass of a Bloch mode determines to great extent the backscattering properties of the system~\cite{Fagiani}.\ In photonic lattices, the backscattering is enhanced by a high density of optical states, as in flat bands or band edges.\ In general, these are the spectral regions where backscattering-induced localization appears for lower amounts of disorder.\ At the band edge of a photonic crystal or at the cutoff frequency of a guided mode it is possible to obtain an effective Schr\"{o}dinger equation where the random potential derives from the structural disorder of the system and the effective kinetic term, $\nu(k)\approx k^2 (\partial^2 \nu/ \partial^2 k)$, presents a minimum value for flat bands thus giving rise to bound states and Anderson localization.\ This is a possible physical picture to understand this process intuitively.\ Based on their values and close to the cutoff frequency as we are in our calculations, we should expect a much stronger backscattering in the valley-Hall edge state than in the conventional waveguide for the same amount of disorder and group index value.

      \end{widetext}


\begin{thebibliography}{99}


\bibitem{Atoms}
L. V. Hau, S. E. Harris, Z. Dutton, C.H. Behroozi. Nature \textbf{397}, 594 (1999).

\bibitem{Monat}
C. Monat et al, Optics Express \textbf{17}, 2944 (2009)

\bibitem{Bajcsy}
M. Bajcsy, S. Hofferberth, V. Balic, T. Peyronel, M. Hafezi, A. S. Zibrov, V. Vuletic, and M. D. Lukin. Phys. Rev. Lett. \textbf{102}, 203902 (2009).

\bibitem{Stenner}
M.D. Stenner et al, Optics Express \textbf{13}, 9995 (2005)

\bibitem{Arcari}
M. Arcari et al, Phys. Rev. Lett. \textbf{113}, 093603 (2014)

\bibitem{Sayrin}
C. Sayrin, C. Clausen, B. Albrecht, P. Schneeweiss, and A. Rauschenbeutel. Optica \textbf{2}, 353 (2015)

\bibitem{Ek}
S. Ek et al, Nature Communications \textbf{5}, 5039 (2014)

\bibitem{polaritons}
K. Jachymski, P. Bienias, and H.P. B\"{u}chler. Phys. Rev. Lett. \textbf{117}, 053601 (2016).

\bibitem{plasmons}
I.Y. Park et al, Nature Photonics \textbf{5}, 677 (2011).

\bibitem{spinors}
J. Ruseckas, V. Kudrias\u{o}v, A. Mekys, T. Andrijauskas, Ite A. Yu, and G. Juzeliunas. Phys. Rev. A \textbf{98}, 013846 (2018).

\bibitem{magnons}
C. Kong et al, Optics Express \textbf{27}, 5544 (2019).

\bibitem{excitons}
N.J. Thompson et al, Appl. Phys. Lett. \textbf{103}, 263302 (2013).

\bibitem{Krauss1}
T.F Krauss. Nature Photonics \textbf{2}, 448 (2008).

\bibitem{Krauss2}
T.F. Krauss et al, Nature \textbf{383}, 699 (1996).

\bibitem{Baba}
T. Baba. Nature Photonics \textbf{2}, 465 (2008).

\bibitem{Mork}
S. Ek, P. Lunnemann, Y. Chen, E. Semenova, K. Yvind , J. Mork. Nature Communications \textbf{5}, 5039 (2014).

\bibitem{Krauss3}
J.F. Priolo, T. Gregorkiewicz, M. Galli, T.F. Krauss. Nature Nanotechnology \textbf{9}, 19 (2014).

\bibitem{Integrated}
J. Wang et al, Nature Photonics \textbf{14}, 273 (2020).

\bibitem{Joannopoulos}
\textit{Photonic Crystals: Molding the Flow of Light}. J.D. Joannopoulos, S.G. Johnson, J.N. Winn, and R.D. Meade. Princeton University Press (2011).

\bibitem{Red}
S.E. Skipetrov and J.H. Page. New J. Phys. \textbf{18} 021001 (2016).

\bibitem{PD1}
P.D. Garc\'{i}a, A. Javadi, H. Thyrrestrup, and P. Lodahl. Appl. Phys. Lett. \textbf{102}, 031101 (2013).

\bibitem{Galli}
Y. Lai, S. Pirotta, G. Urbinati, D. Gerace, M. Minkov, V. Savona, A. Badolato, and M. Galli. Appl. Phys. Lett. \textbf{104}, 241101 (2014).

\bibitem{Topolancik}
J. Topolancik, B. Ilic, and F. Vollmer. Phys. Rev. Lett. \textbf{99}, 253901 (2007).

\bibitem{Anderson}
P.~W. Anderson. Phys. Rev. \textbf{109}, 1492 (1958).

\bibitem{TopologicalPhotonics}
T. Ozawa et al, Rev. Mod. Phys. \textbf{91}, 015006 (2019).

\bibitem{AQHE}
S. Raghu and F.D.M. Haldane. Phys. Rev. A \textbf{78}, 033834 (2008).

\bibitem{Marin}
Z. Wang, Y. Chong, J. D. Joannopoulos, M. Solja\v{c}i\'{c}. Nature \textbf{461}, 772 (2009).

\bibitem{QSHE}
L.H. Wu and X. Hu. Phys. Rev. Lett. \textbf{114}, 223901 (2015).

\bibitem{valley}
T. Ma and G. Shvets. New J. Phys. \textbf{18}, 025012 (2016).

\bibitem{valley_exp}
X.T. He, et al, Nature Communications \textbf{10}, 872 (2019).

\bibitem{Mazoyer}
S. Mazoyer, J. P. Hugonin, and P. Lalanne. Phys. Rev. Lett. \textbf{103}, 063903 (2009).

\bibitem{Sauer}
E. Sauer, J. P. Vasco, S. Hughes. Preprint can be found here: arXiv:2005.12828v1 (2020).

\bibitem{EngineeredSlowLight}
T. Baba and D. Mori. Journal of Physics D: Applied Physics \textbf{40}, 9 (2007).

\bibitem{BZWinding}
J. Guglielmon and M. C. Rechtsman. Phys. Rev. Lett. \textbf{122}, 153904 (2019).

\bibitem{PD3}
P.D. Garc\'{i}a, S. Smolka, S. Stobbe, and P. Lodahl. Phys. Rev. B \textbf{82}, 165103 (2010).

\bibitem{PD4}
P.D. Garc\'{i}a, G. Kir\v{s}ansk\.{e}, A. Javadi, S. Stobbe, and P. Lodahl. Phys. Rev. B \textbf{96}, 144201 (2017).

\bibitem{Garcia-Martin}
A. Garc\'{i}a-Mart\'{i}n and J. J. S\'{a}enz. Phys. Rev. Lett. \textbf{87} (11), 116603 (2001).

\bibitem{Froufe}
L. S. Froufe-P\'{e}rez, P. Garc\'{i}a-Mochales, P. A. Serena, P. A. Mello, and J. J. S\'{a}enz. Phys. Rev. Lett. \textbf{89}, 246403 (2002).

\bibitem{MacKinnon}
A. MacKinnon and B. Kramer. Phys. Rev. Lett. \textbf{47}, 1546 (1981).

\bibitem{Savona}
V. Savona. Phys. Rev. B \textbf{83}, 085301 (2011).

\bibitem{Fleury}
B. Orazbayev, R. Fleury. Nanophotonics \textbf{8}, 8 2019.

\bibitem{Sheng}
\textit{Introduction to Wave Scattering, Localization, and Mesoscopic Phenomena}. P. Sheng (Academic Press, San Diego (1995)).

\bibitem{Smolka}
S. Smolka et al, New J. Phys. \textbf{13}, 063044 (2011).

\bibitem{Fagiani}
R. Faggiani et al, Scientific Reports \textbf{6}, 27037 (2016).

\bibitem{AQHE_exp}
Z. Wang, Y. D. Chong, J. D. Joannopoulos, and M. Solja\v{c}i\'{c}. Phys. Rev. Lett. \textbf{100}, 013905 (2008).

\bibitem{QSHE_exp}
S. Barik, et al, Science \textbf{359}, 666 (2018).

\bibitem{Shalaev}
M.I. Shalaev, et al, Nature Nanotechnology \textbf{14}, 31 (2019).

\bibitem{Goban}
A. Goban et al, Nat Commun. \textbf{5}, 3808 (2014).

\bibitem{Aitzol}
M. Blanco de Paz et al, Advanced Quantum Technologies \textbf{3}, 1900117 (2020).


\scriptsize


\end{thebibliography}

\begin{thebibliography}{11}

\bibitem{WuHu}
H. Wu and X. Hu. \emph{Scheme for achieving a topological photonic crystal by using dielectric material}. Phys. Rev. Lett. \textbf{114}, 223901 (2015).

\bibitem{Benakker}
A. Rycerz, J. Tworzyd\l{}o, C. Beenakker. \emph{Valley filter and valley valve in graphene}. Nat. Phys. \textbf{3}, 172 (2007).

\bibitem{magnetic}
D. Xiao and N. YaoWand. \emph{Valley-contrasting physics in graphene: magnetic moment and topological transport.} Phys. Rev. Lett. \textbf{99},236809 (2007).

\bibitem{valley}
T. Ma and G. Shvets. \emph{All-Si Valley-Hall photonic topological insulator}. New J. Phys. \textbf{18}, 025012 (2016).

\bibitem{Mazoyer}
S. Mazoyer, J. P. Hugonin, and P. Lalanne. Phys. Rev. Lett. \textbf{103}, 063903 (2009).

\bibitem{Fagiani}
R. Faggiani et al, Scientific Reports \textbf{6}, 27037 (2016).


\end{thebibliography}
\end{document}